\documentclass[conference,10pt]{IEEEtran}

\usepackage[utf8]{inputenc}
\usepackage[T1]{fontenc}
\usepackage{amsmath,amssymb,amsfonts}
\usepackage{graphicx}
\usepackage{textcomp}
\usepackage{xcolor}
\usepackage{tikz}
\usepackage{pgfplots}
\pgfplotsset{compat=1.18}
\usetikzlibrary{shapes.geometric, arrows.meta, positioning, fit, calc, backgrounds, decorations.markings, patterns}
\usepackage{booktabs}
\usepackage{multirow}
\usepackage{tabularx}
\usepackage{colortbl}
\usepackage{hyperref}
\usepackage{cleveref}
\usepackage{enumitem}
\usepackage{balance}
\usepackage{float}
\usepackage{cite}
\usepackage{listings}
\usepackage{comment}
\usepackage[table]{xcolor}
\usepackage{array}
\usepackage{tabularx}
\usepackage[table]{xcolor}

\usepackage{tikz}
\usetikzlibrary{arrows.meta,positioning}
\usepackage{graphicx}

\definecolor{colorgray}{gray}{0.92}
\definecolor{darkblue}{RGB}{0,51,102}
\definecolor{midblue}{RGB}{70,130,180}
\definecolor{darkred}{RGB}{160,40,40}
\definecolor{darkgreen}{RGB}{0,100,60}
\definecolor{rowgray}{RGB}{240,240,245}

\lstdefinelanguage{dapl}{
    basicstyle=\ttfamily\scriptsize,
    keywordstyle=\color{darkblue}\bfseries,
    commentstyle=\color{darkgreen!70!black}\itshape,
    stringstyle=\color{darkred!80!black},
    keywords={actor_id, subject_token, tenant_id, roles, enforcement_mode, policy_ids, correlation_id, statement_id, cursor_id, read_only, max_rows, max_bytes, timeout_ms, schema, payload, lineage, resources, rows, bytes, elapsed_ms, true, false, null},
    morestring=[b]",
    morecomment=[l]{//},
    showstringspaces=false,
    breaklines=true,
    frame=single,
    framerule=0.3pt,
    rulecolor=\color{darkblue!40},
    backgroundcolor=\color{darkblue!3},
    tabsize=2,
    columns=fullflexible
}

\IEEEoverridecommandlockouts
\IEEEaftertitletext{\vspace{-1\baselineskip}}

\title{GATE-DT: Trust-Gated Operational Digital Twins for Secure Distributed Retail and E-Commerce}

\author{
\IEEEauthorblockN{\makebox[0.24\textwidth][c]{1\textsuperscript{st} Pradnil Uday Rajankar}}
\IEEEauthorblockA{
\makebox[0.20\textwidth][c]{\textit{IEEE Senior Member}} \\
\makebox[0.20\textwidth][c]{Independent Researcher} \\
\makebox[0.20\textwidth][c]{Seattle, Washington, USA} \\
\makebox[0.20\textwidth][c]{pradnil.rajankar@yahoo.com}}
\and

\IEEEauthorblockN{\makebox[0.24\textwidth][c]{2\textsuperscript{nd} Rohit Singh}}
\IEEEauthorblockA{
\makebox[0.20\textwidth][c]{\textit{IEEE Member}} \\
\makebox[0.20\textwidth][c]{George Washington University} \\
\makebox[0.20\textwidth][c]{Washington, DC, USA} \\
\makebox[0.20\textwidth][c]{rohit.singh1@gwu.edu}}
\and

\IEEEauthorblockN{\makebox[0.24\textwidth][c]{3\textsuperscript{rd} Kshitij Sharma}}
\IEEEauthorblockA{
\makebox[0.20\textwidth][c]{\textit{IEEE Senior Member}} \\
\makebox[0.20\textwidth][c]{Independent Researcher} \\
\makebox[0.20\textwidth][c]{Cupertino, California, USA} \\
\makebox[0.20\textwidth][c]{kshitijsharma@ieee.org}}
\and [\hfill\mbox{}\par\mbox{}\hfill]

\IEEEauthorblockN{\makebox[0.24\textwidth][c]{4\textsuperscript{th} Naresh Alapati}}
\IEEEauthorblockA{
\makebox[0.20\textwidth][c]{\textit{IEEE Senior Member}} \\
\makebox[0.20\textwidth][c]{Independent Researcher} \\
\makebox[0.20\textwidth][c]{Bentonville, Arkansas, USA} \\
\makebox[0.20\textwidth][c]{narencachy@gmail.com}}
\and

\IEEEauthorblockN{\makebox[0.24\textwidth][c]{5\textsuperscript{th} Koteswararao Nallabothu}}
\IEEEauthorblockA{
\makebox[0.20\textwidth][c]{\textit{IEEE Member}} \\
\makebox[0.20\textwidth][c]{Independent Researcher} \\
\makebox[0.20\textwidth][c]{Charlotte, North Carolina, USA} \\
\makebox[0.20\textwidth][c]{nallabothukoteswar@gmail.com}}
\and

\IEEEauthorblockN{\makebox[0.24\textwidth][c]{6\textsuperscript{th} Ankit Talwar}}
\IEEEauthorblockA{
\makebox[0.20\textwidth][c]{Independent Researcher} \\
\makebox[0.20\textwidth][c]{Austin, Texas, USA} \\
\makebox[0.20\textwidth][c]{ankittalwar@gmail.com}}
\and [\hfill\mbox{}\par\mbox{}\hfill]

\IEEEauthorblockN{\makebox[0.24\textwidth][c]{7\textsuperscript{th} Manjari Mettu}}
\IEEEauthorblockA{
\makebox[0.20\textwidth][c]{\textit{IEEE Member}} \\
\makebox[0.20\textwidth][c]{Independent Researcher} \\
\makebox[0.20\textwidth][c]{Chandler, Arizona, USA} \\
\makebox[0.20\textwidth][c]{manjari2878@gmail.com}}
\and

\IEEEauthorblockN{\makebox[0.24\textwidth][c]{8\textsuperscript{th} Surabhi Agarwal}}
\IEEEauthorblockA{
\makebox[0.20\textwidth][c]{Independent Researcher} \\
\makebox[0.20\textwidth][c]{Sunnyvale, California, USA} \\
\makebox[0.20\textwidth][c]{surabhia@tepper.cmu.edu}}
\and

\IEEEauthorblockN{\makebox[0.24\textwidth][c]{9\textsuperscript{th} Joyjit Roy}}
\IEEEauthorblockA{
\makebox[0.20\textwidth][c]{\textit{IEEE Senior Member}} \\
\makebox[0.20\textwidth][c]{Independent Researcher} \\
\makebox[0.20\textwidth][c]{Austin, Texas, USA} \\
\makebox[0.20\textwidth][c]{joyjit.roy.tech@gmail.com}}
\vspace*{-30pt}
}

\begin{document}

\maketitle

\begin{abstract}
Autonomous agents increasingly support retail and e-commerce operations, yet their decisions may rely on stale, inconsistent, or compromised distributed state. This paper presents Governed Autonomous Transaction Execution for Digital Twins (GATE-DT), a trust-gated operational digital-twin framework that separates autonomous decision-making from execution authority. GATE-DT evaluates six trust dimensions: data quality, operational-twin fidelity, model confidence, distributed-state reliability, cybersecurity integrity, and decision provenance. A deterministic authorization plane enforces trust thresholds and hard vetoes, while adaptive autonomy restricts execution under trust degradation and progressively restores authority during recovery. Evaluation with a distributed commerce emulation harness covers state inconsistencies, infrastructure failures, cyber and fraud attacks, demand surges, and compound disruptions. Under compound stress, GATE-DT reduced untrustworthy autonomous execution from 8.8\% to 1.9\% (78.4\%, $p < 0.001$). Stale-state execution declined from 5.9\% to 0.9\%, fraud acceptance from 5.3\% to 1.4\%, and overselling from 3.0\% to 0.8\%. These improvements came with an additional 19~ms of p95 authorization latency and an increase in legitimate holds from 0.9\% to 4.1\%, illustrating the trade-off between execution safety and operational continuity.
\end{abstract}


\section{Introduction}
\label{sec:intro}

\textbf{Autonomous Commerce.}
Retail and e-commerce platforms are evolving from conventional transaction processing toward autonomous operations. Autonomous agents support pricing, procurement, inventory coordination, fulfillment planning, and fraud analysis. Recent research demonstrates privacy-sensitive business-to-business (B2B) negotiation agents operating under privacy, policy, and auditability constraints \cite{roy2026negotiation}. Digital twins provide synchronized representations of logistics, supply chains, and operational processes to support decision-making \cite{le2024digitaltwins,ashrafian2023digitaltwin}. ShoppingBench \cite{wang2026shoppingbench} demonstrates autonomous interaction with realistic e-commerce tasks, while WebArena \cite{zhou2024webarena} highlights challenges in executing tasks across functional websites. These advances underscore the need to establish whether the operational state enabling autonomous actions is sufficiently trustworthy before execution.

\textbf{Distributed-State Reliability.}
Modern commerce platforms distribute inventory, reservations, checkout, payments, orders, and fulfillment across replicated databases, event streams, caches, and services. Distributed storage balances availability and consistency during failures and network partitions \cite{decandia2007dynamo}, while systems such as Spanner provide stronger consistency guarantees \cite{corbett2012spanner}. However, availability and consistency do not establish whether observed state is current and reliable for an autonomous action. Replication lag, stale projections, consumer lag, reordered or duplicate events, and state divergence can undermine decision contexts. Reliability requirements are action-dependent: bounded-stale state may suffice for discovery or planning, whereas inventory reservations, payments, and order transitions demand stronger guarantees.

\textbf{Cybersecurity and Execution.}
Cybersecurity adds further uncertainty. Seemingly valid actions may depend on a state compromised by stolen credentials, malicious API activity, replayed events, forged messages, manipulated transaction data, or compromised services. Zero-trust architectures emphasize explicit authentication and authorization of users, services, and applications, rather than relying on network location or ownership \cite{nist2020zerotrust}. Similar concerns arise when agentic systems generate actions with operational consequences. For example, AgentSOC combines generative reasoning with feasibility validation, risk-aware action selection, and policy and safety guardrails before executing security responses \cite{roy2026agentsoc}. The distinction between generating plausible actions and ensuring safe, authorized execution is critical for autonomous commerce, where authorization must also consider the integrity, freshness, consistency, and provenance of the distributed transactional state supporting an action.

\textbf{Sustainable Development Goals.}
This work aligns with three United Nations Sustainable Development Goals (SDGs). It supports SDG~9 (Industry, Innovation, and Infrastructure) by improving the resilience and security of distributed digital commerce infrastructure. It supports SDG~8 (Decent Work and Economic Growth) by enabling safe autonomous operations that sustain transaction continuity for retail businesses. It supports SDG~12 (Responsible Consumption and Production) by reducing overselling, duplicate orders, and fraudulent transactions that cause waste and supply-chain inefficiency.
\section{Background and Related Work}
\label{sec:background}

\subsection{Digital Twins and Autonomous Retail Operations}

Digital twins provide synchronized models for monitoring, analysis, and decision support. Research on logistics and supply-chain digital twins \cite{le2024digitaltwins} and automated order picking and fulfillment \cite{ashrafian2023digitaltwin} demonstrates their relevance to commerce. Ivanov and Dolgui \cite{ivanov2021supplytwin} examine digital supply-chain twins for managing disruption risks and improving resilience. Privacy-preserving negotiation agents \cite{roy2026negotiation} support autonomous business interactions, while ReAct \cite{yao2023react} combines reasoning with environment-facing actions. ShoppingBench \cite{wang2026shoppingbench} and WebArena \cite{zhou2024webarena} benchmark shopping and web-task agents. Although these works advance operational modeling, autonomous reasoning, and task execution, they lack a unified mechanism for authorizing agent-proposed commerce actions based on the freshness, consistency, and security of distributed transactional state.

\subsection{Distributed Transaction Reliability and State Freshness}

Distributed storage systems provide varying consistency and availability guarantees. Dynamo \cite{decandia2007dynamo} prioritizes availability through replication, object versioning, and application-assisted conflict resolution, accepting weaker consistency during failures. Spanner \cite{corbett2012spanner} provides externally consistent distributed transactions, while probabilistically bounded staleness \cite{bailis2012pbs} characterizes freshness and latency trade-offs under partial-quorum replication. The saga model \cite{garciamolina1987sagas} introduces compensating transactions to manage partial completion in long-running transactional workflows. The Dataflow Model \cite{akidau2015dataflow} examines correctness and latency trade-offs in processing unbounded, out-of-order event streams, while age-of-information research \cite{yates2021age} quantifies the freshness of observed state. In retail, services may expose stale inventory, payment, or order data, and recent updates can remain inconsistent across systems. Although these approaches address consistency, freshness, event processing, and transaction recovery, they do not directly integrate distributed-state reliability with cybersecurity integrity to authorize individual autonomous actions.

\subsection{Cybersecurity, Zero Trust, and Runtime Authorization}

Zero-trust architectures require explicit authorization across identities, services, and applications rather than relying on network location \cite{nist2020zerotrust}. Zanzibar \cite{pang2019zanzibar} provides consistent authorization across globally distributed services, while Cedar \cite{cutler2024cedar} introduces an expressive and analyzable policy language for evaluating authorization requests. AgentSOC \cite{roy2026agentsoc} incorporates feasibility validation, risk-aware action selection, and policy safeguards into agentic security operations. AgentDojo~\cite{debenedetti2024agentdojo} evaluates prompt-injection attacks and defenses for tool-using agents, while Progent~\cite{shi2025progent} provides programmable privilege control for LLM agent tool calls. However, these approaches do not directly assess the trustworthiness of distributed commerce state during transaction authorization. Authenticated agents may still process replayed inventory events, inconsistent order records, or unresolved payment outcomes. These limitations motivate runtime authorization that jointly evaluates cybersecurity integrity, distributed-state reliability, operational fidelity, and decision provenance before allowing transaction-sensitive autonomous actions. Table~\ref{tab:related_work} summarizes this separation, which GATE-DT addresses through trust-gated execution.

\begin{table}[t]
\caption{Comparison of Related Approaches and GATE-DT}
\label{tab:related_work}
\centering
\footnotesize
\setlength{\tabcolsep}{2pt}
\renewcommand{\arraystretch}{1.12}
\arrayrulecolor{black}
\setlength{\arrayrulewidth}{0.8pt}

\begin{tabular*}{\columnwidth}{
@{\extracolsep{\fill}}
>{\columncolor{colorgray}}p{0.58\columnwidth}
>{\columncolor{white}}c
>{\columncolor{colorgray}}c
>{\columncolor{white}}c
>{\columncolor{colorgray}}c
>{\columncolor{white}}c
@{}
}
\hline

\textbf{Approach} &
\textbf{DT} &
\textbf{Agent} &
\textbf{Dist.} &
\textbf{Cyber} &
\textbf{Auth.} \\

\hline

Logistics/Supply-Chain DTs
\cite{le2024digitaltwins,ashrafian2023digitaltwin,ivanov2021supplytwin}
& $\checkmark$ & -- & $\circ$ & -- & -- \\

Autonomous Web/Shopping Agents
\cite{wang2026shoppingbench,zhou2024webarena,yao2023react}
& -- & $\checkmark$ & -- & -- & -- \\

Distributed Transactions/Event Processing
\cite{corbett2012spanner,garciamolina1987sagas,akidau2015dataflow}
& -- & -- & $\checkmark$ & -- & -- \\

Zero-Trust Architecture
\cite{nist2020zerotrust}
& -- & -- & -- & $\checkmark$ & $\checkmark$ \\

Distributed Authorization Systems
\cite{pang2019zanzibar,cutler2024cedar}
& -- & -- & $\circ$ & $\circ$ & $\checkmark$ \\

LLM Agent Privilege Control
\cite{shi2025progent}
& -- & $\checkmark$ & -- & $\circ$ & $\checkmark$ \\

\textbf{GATE-DT}
& $\checkmark$ & $\checkmark$ & $\checkmark$
& $\checkmark$ & $\checkmark$ \\

\hline
\end{tabular*}

\vspace{1mm}
\parbox{\columnwidth}{
\scriptsize
\textit{DT}: operational digital-twin state;
\textit{Agent}: autonomous decision support;
\textit{Dist.}: distributed-state reliability;
\textit{Cyber}: cybersecurity evidence;
\textit{Auth.}: runtime authorization.
$\checkmark$ = explicit support;
$\circ$ = partial/indirect support;
-- = not a primary focus.
}
\end{table}

\section{Framework and System Architecture}
\label{sec:arch}

\subsection{System Model and Commerce Operations Digital Twin}

\begin{figure}[t]
    \centering
    \includegraphics[width=\columnwidth]{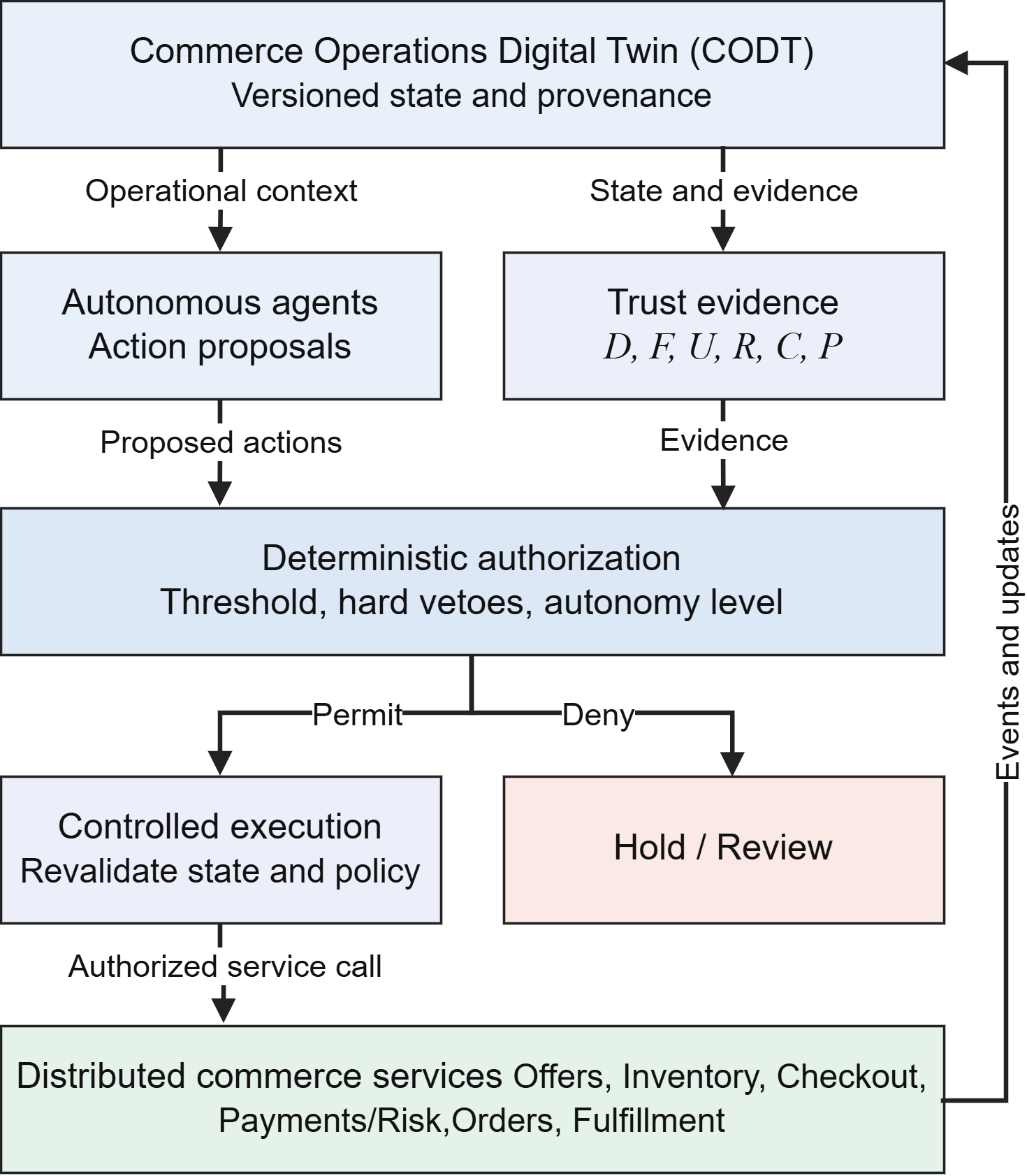}
    \caption{GATE-DT system architecture showing the Commerce Operations Digital Twin, autonomous action proposals, trust evidence, deterministic authorization, and controlled execution.}
    \label{fig:architecture}
\end{figure}

GATE-DT models a distributed retail and e-commerce platform comprising offer, inventory, reservation, checkout, payment, fraud, order, and fulfillment services. These services exchange transactional state through replicated databases, event streams, and service interfaces. The Commerce Operations Digital Twin (CODT) maintains a synchronized, versioned representation of these workflows based on service updates and transaction events. Each state element retains source, timestamp, and version metadata to support freshness, consistency, and provenance assessment. Since updates can be delayed, duplicated, reordered, or compromised, the CODT may diverge from the underlying platform. Autonomous agents can propose operational actions but cannot execute them directly. A deterministic authorization plane evaluates each proposal against trust evidence and action-specific constraints before permitting execution (Fig.~\ref{fig:architecture}).

\noindent\textbf{Threat Model and Trust Assumptions.} The adversary may submit malicious requests, replay or forge events, misuse credentials, and delay, reorder, or manipulate distributed updates. These attacks may affect CODT projections, provenance metadata, and the evidence used to assess cybersecurity integrity ($C$) and provenance ($P$). The CODT and agent-generated proposals are not considered authoritative. Execution decisions require validation against authoritative service state and security evidence. The security model assumes the authorization plane, its policy configuration, enforcement points, and mechanisms for verifying authoritative evidence remain trustworthy. Compromise of these components or their trust anchors is outside the security guarantees of GATE-DT.

\noindent\textbf{Authorization-Plane Availability.} When required evidence or quorum status cannot be verified, autonomous execution must be denied or paused, rather than authorized based on outdated CODT state or prior decisions. Direct compromise or complete unavailability of the authorization plane is not separately evaluated. Resilience to these failures is outside the demonstrated evaluation scope.

\subsection{Trust Evidence and Quantification}

For each agent-proposed action $a$ at time $t$, GATE-DT generates a six-dimensional trust-evidence vector $\boldsymbol{z}(a,t)=[D,F,U,R,C,P]^\top$, representing data quality, operational-twin fidelity, model confidence, distributed-state reliability, cybersecurity integrity, and decision provenance. Each dimension is normalized to $[0,1]$ based on action-specific evidence.

Data quality $D$ is the fraction of valid schema and field checks. Fidelity $F$ measures agreement with the authoritative service state. Model confidence $U$ is computed as $U=\min(1,\max(0,c_a))$, where $c_a$ is the action-level confidence value returned by the LLM-based agent. This score represents the agent's reported confidence rather than an empirically calibrated probability. Reliability $R$ combines freshness, version consistency, and quorum validity. Freshness scoring is independent of the hard-veto threshold. Cybersecurity integrity $C$ assesses authentication, message integrity, and replay protection. Provenance $P$ evaluates the completeness and verifiability of evidence. Each check-based score reflects the fraction of applicable checks satisfied. Missing or unverifiable required evidence assigns zero to the affected dimension. Aggregate trust is computed as

\begin{equation}
T(a,t)=\prod_{j\in\mathcal{J}}z_j(a,t)^{w_j},
\quad \sum_{j\in\mathcal{J}}w_j=1,
\label{eq:trust}
\end{equation}

Here, $\mathcal{J}=\{D,F,U,R,C,P\}$ and $w_j>0$ denotes the weight of dimension $j$. Table~\ref{tab:trust_policy} outlines the reference trust policy, which prioritizes cybersecurity integrity and distributed-state reliability. The thresholds $\delta_j$ set the minimum acceptable evidence quality and serve as policy constraints, not empirically calibrated cutoffs. The weights $w_j$ define each dimension's influence on aggregate trust and are constant across scenarios. The sensitivity analysis varies the threshold $\tau_a$, the dimension minimums $\delta_j$, and the maximum critical-state age independently. This multiplicative formulation penalizes weak evidence and yields zero trust if any dimension is zero. Such actions also fail the corresponding minimum $\delta_j$ in \eqref{eq:authorization}, so zero collapse acts as a fail-closed condition rather than an additional source of denial.

\begin{table}[t]
\caption{Reference Trust Policy}
\label{tab:trust_policy}
\centering
\footnotesize
\setlength{\tabcolsep}{4pt}
\renewcommand{\arraystretch}{1.12}
\arrayrulecolor{black}
\setlength{\arrayrulewidth}{0.8pt}

\begin{tabular}{@{}c@{}}
\begin{tabular}{
@{}
>{\columncolor{colorgray}}p{0.07\columnwidth}
@{\hspace{6pt}}
>{\columncolor{colorgray}}p{0.45\columnwidth}
>{\columncolor{white}\centering\arraybackslash}p{0.09\columnwidth}
>{\columncolor{white}\centering\arraybackslash}p{0.09\columnwidth}
@{}}
\hline
\textbf{Dim.} &
\textbf{Evidence} &
$\boldsymbol{w_j}$ &
$\boldsymbol{\delta_j}$ \\
\hline
$D$ & Schema and field validity & 0.10 & 0.62 \\
$F$ & Agreement with service state & 0.10 & 0.67 \\
$U$ & Agent-reported confidence & 0.10 & 0.54 \\
$R$ & Freshness, versions, quorum & 0.25 & 0.74 \\
$C$ & Authentication, integrity, replay & 0.30 & 0.82 \\
$P$ & Evidence completeness & 0.15 & 0.63 \\
\hline
\end{tabular}
\\[1mm]
{\scriptsize
\textit{A3 reference policy}: $\tau_a=0.80$;
maximum critical-state age = 0.5~s.}
\end{tabular}

\end{table}

\subsection{Deterministic Execution Authorization}

Runtime verification provides a foundation for monitoring system behavior against specified correctness and safety properties \cite{bartocci2018runtime}. GATE-DT applies this principle by separating agent-generated decisions from execution authority and evaluating runtime trust evidence before permitting transaction-sensitive actions. For each action $a$, the authorization plane evaluates aggregate trust $T(a,t)$ against an action-specific threshold $\tau_a$, dimension-specific minimum scores $\delta_j$, a hard-veto indicator $V(a,t)$, and the autonomy level $L(t)$. Let $\ell(a)$ denote the minimum autonomy level required for execution. Authorization is defined as

\begin{equation}
\operatorname{Permit}(a,t)=
\begin{cases}
1, &
\begin{aligned}
&T(a,t)\geq\tau_a,\quad V(a,t)=0,\\
&z_j(a,t)\geq\delta_j,\quad \forall j\in J,\\
&L(t)\geq\ell(a),
\end{aligned}\\
0, & \text{otherwise}.
\end{cases}
\label{eq:authorization}
\end{equation}

\begin{figure}[t]
    \centering
    \includegraphics[width=\columnwidth]{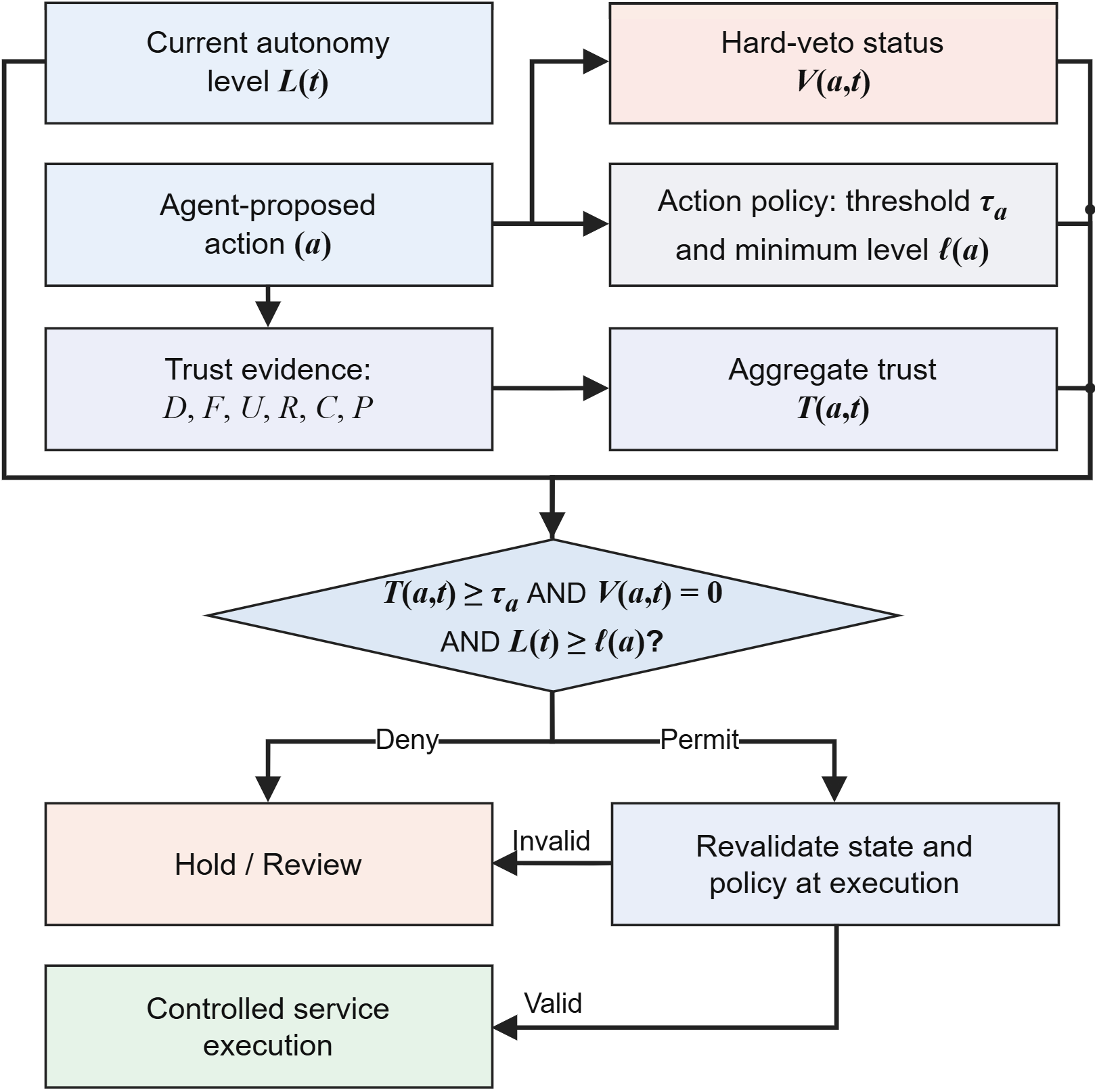}
    \caption{GATE-DT trust-gated execution authorization workflow showing trust evaluation, hard-veto enforcement, autonomy-level verification, and controlled execution. Authorization additionally requires each trust dimension to satisfy $z_j(a,t)\geq\delta_j$ for all $j\in\mathcal{J}$.}
    \label{fig:trust_authorization}
\end{figure}

Authorization requires meeting both the aggregate trust threshold and the dimension-specific minimums $\delta_j$. Strong evidence in one dimension cannot compensate for weak evidence in another. Critical security and transactional violations trigger separate, action-specific hard vetoes, indicated by $V(a,t)=1$. These violations include invalid authentication, replayed events, inventory-version conflicts, unresolved payment outcomes, and missing required quorums. Denied actions are held or referred for review. Fig.~\ref{fig:trust_authorization} illustrates the authorization workflow.

Permitted actions require version-bound execution preconditions. Each affected service atomically validates relevant state versions and policy constraints during its local transaction commit. Any mismatch triggers rejection and reauthorization. Multi-service workflows repeat these checks at each transactional boundary. If a workflow partially commits and a later operation fails, GATE-DT suspends autonomous execution and initiates compensating transactions for reversible operations. Unresolved or irreversible outcomes must be reconciled before execution authority is restored. Each logical operation uses an idempotency key that remains unchanged across retries, and services enforce durable deduplication on this key to prevent repeated payment, reservation, or order side effects.

\subsection{Adaptive Autonomy and Recovery}

\begin{figure*}[t]
    \centering
    \includegraphics[width=\textwidth]
    {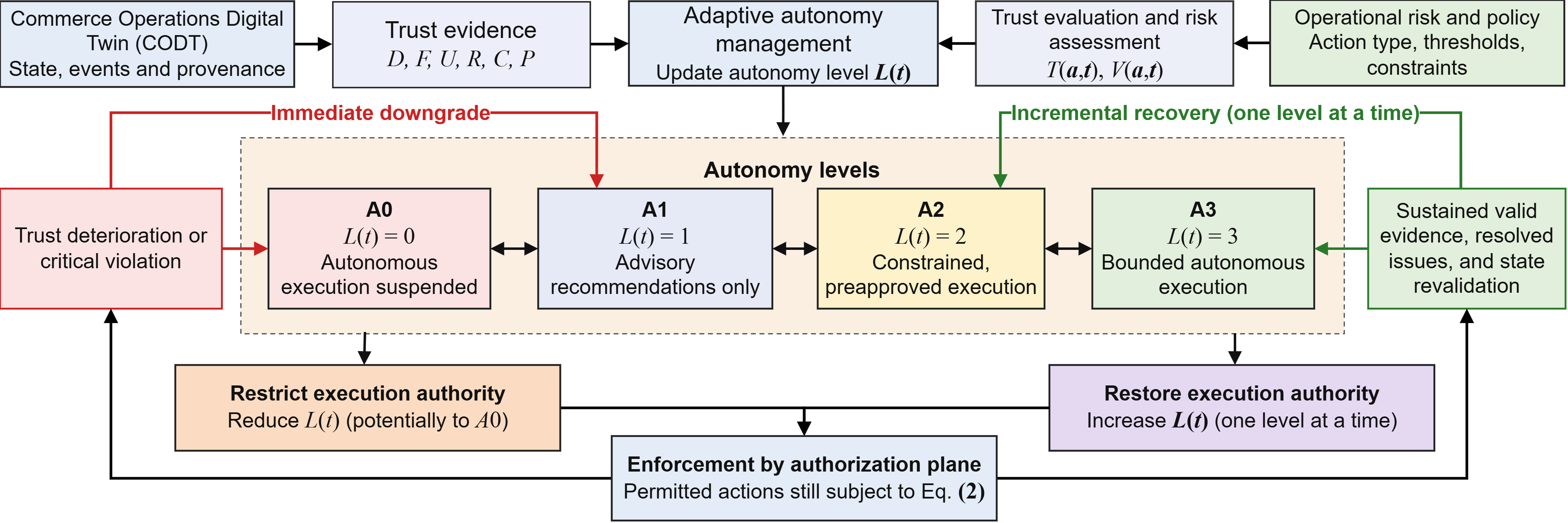}
    \caption{GATE-DT adaptive autonomy and recovery framework showing four autonomy levels (A0--A3), immediate restriction under trust deterioration, incremental recovery, and authorization-plane enforcement.}
    \label{fig:adaptive_autonomy}
\end{figure*}

Runtime assurance architectures such as Black-Box Simplex demonstrate how monitored safety conditions can trigger a transfer of control from unsafe autonomous components \cite{mehmood2022simplex}. Building on this principle, GATE-DT dynamically adjusts the autonomy level $L(t) \in \{0,1,2,3\}$ according to trust evidence and operational risk. At A0, all autonomous actions are suspended. A1 allows only advisory recommendations, without execution. A2 permits constrained, preapproved low-risk actions. A3 grants bounded autonomy under action-specific policies. Executable autonomous actions require $\ell(a)\in\{2,3\}$. If trust declines or a critical security violation occurs, execution authority is immediately restricted, potentially reducing autonomy to A0. Recovery requires sustained valid evidence, resolution of blocking conditions, and revalidation of the affected transactional state. Autonomy is restored incrementally, one level at a time, to prevent rapid oscillation. Permitted execution remains subject to (\ref{eq:authorization}), while Fig.~\ref{fig:adaptive_autonomy} illustrates adaptive autonomy and recovery.

\section{Evaluation Methodology}
\label{sec:evaluation}

\subsection{Evaluation Design and Baselines}

The evaluation examines five configurations: B0 (centralized transaction processing), B1 (distributed processing without trust gating), B2 (distributed processing with conventional security controls), B3 (agentic execution under conventional security controls without trust gating), and GATE-DT (agentic execution with trust evidence, hard vetoes, and adaptive autonomy). Comparable workloads and disruption scenarios are applied across configurations. B3 and GATE-DT share the same distributed infrastructure, security controls, and agent proposals. Their comparison evaluates the combined effect of GATE-DT's trust-gated authorization and transactional execution safeguards, rather than attributing all improvements to trust gating alone.

A Meta Llama 3.1 8B Instruct agent generates action proposals using transaction context and distributed operational-state information. These proposals include inventory adjustments, reservations, payment operations, fraud handling, order-state transitions, and fulfillment decisions. The agent-reported confidence dimension $U$ contributes to the trust assessment for execution authorization.

\subsection{Workload and System Configuration}

The evaluation uses a parameterized emulation harness representing geographically distributed retail operations. As detailed in Table~\ref{tab:eval_config}, the configuration comprises three regions, nine database nodes, six event brokers, 96 event partitions, and 32 fulfillment locations supporting 500,000 stock-keeping units (SKUs). These parameters define the logical infrastructure topology, not a specific commercial database, messaging platform, or physical deployment. The workload encompasses inventory updates, reservations, payment events, fraud assessments, and order-state transitions.

Distributed configurations use identical infrastructure parameters, while B0 employs centralized transaction processing. Agent proposals, transaction sequences, and disruption schedules remain consistent across comparable configurations.

\begin{table}[t]
\caption{Evaluation Configuration}
\label{tab:eval_config}
\centering
\footnotesize
\setlength{\tabcolsep}{4pt}
\renewcommand{\arraystretch}{1.12}
\arrayrulecolor{black}
\setlength{\arrayrulewidth}{0.8pt}

\begin{tabular}{
@{}
>{\columncolor{colorgray}}p{0.40\columnwidth}
>{\columncolor{white}}p{0.30\columnwidth}
@{}
}
\hline
\textbf{Parameter} & \textbf{Configuration} \\
\hline

Geographic regions & 3 \\
Fulfillment locations & 32 \\
Product catalog & 500,000 SKUs \\
Distributed database nodes & 9 (3 per region) \\
Event brokers & 6 (2 per region) \\
Event partitions & 96 \\
Normal checkout workload & 50 transactions/s \\
Peak checkout workload & 250 transactions/s \\
Normal event workload & 2,000 events/s \\
Peak event workload & 10,000 events/s \\
Scenario duration & 30 minutes \\
Independent random seeds & 20 per scenario \\

\hline
\end{tabular}
\end{table}

\subsection{Fault, Attack, and Stress Scenarios}

Table~\ref{tab:eval_scenarios} defines six scenarios, from nominal operation (S0) to compound disruption (S5). Each scenario runs for 30 minutes. For S1--S5, disruptions begin at minute 10 and are cleared at minute 20, allowing 10 minutes to observe recovery. Equivalent disruptions are applied where architectural differences permit.


\begin{table}[t]
\caption{Evaluation Scenarios and Injected Disruptions}
\label{tab:eval_scenarios}
\centering
\footnotesize
\setlength{\tabcolsep}{4pt}
\renewcommand{\arraystretch}{1.12}
\arrayrulecolor{black}
\setlength{\arrayrulewidth}{0.8pt}

\resizebox{\columnwidth}{!}{%
\begin{tabular}{
@{}
>{\columncolor{colorgray}}p{0.20\columnwidth}
>{\columncolor{white}}p{0.85\columnwidth}
@{}
}
\hline
\textbf{Scenario} & \textbf{Injected Disruption} \\
\hline
S0: Normal &
Nominal transaction processing and event delivery \\
S1: State &
Replication lag (2--10 s), stale projections, and reordered events \\
S2: Failure &
Service and broker outages, partition disruption, and quorum loss \\
S3: Cyber &
Replayed events, forged updates, invalid credentials, and fraudulent requests \\
S4: Demand &
Fivefold traffic surge, elevated event volume, and inventory contention \\
S5: Compound &
Concurrent replication lag, infrastructure failure, replay attacks, fraudulent requests, and demand surge \\
\hline
\end{tabular}%
}
\end{table}

\subsection{Evaluation Metrics and Protocol}

The evaluation measures execution safety, operational continuity, and authorization overhead.

\textbf{Execution Safety.} Safety is evaluated using the untrustworthy autonomous execution rate (UAER), stale-state execution rate, duplicate side-effect rate, fraud acceptance rate, and overselling incidence. UAER measures the proportion of executed autonomous actions that violate ground-truth consistency, security, or transactional constraints. Stale-state execution involves actions performed with outdated state. Duplicate side effects are repeated payments, reservations, or orders caused by retries or replayed events. Fraud acceptance is the proportion of injected fraudulent requests accepted at the transaction boundary, using identical attack inputs and acceptance criteria across all five configurations.

\textbf{Operational Continuity.} Since trust gating limits the number of actions allowed, UAER is evaluated with the legitimate action hold rate and order completion rate. Legitimate holds are actions blocked despite satisfying ground-truth safety conditions. Recovery time is the interval between disruption removal and restoration of A3 execution authority.

\textbf{Authorization Overhead.} Decision latency is reported as the 95th-percentile (p95) latency for each configuration. Authority-downgrade latency is the time from detection of a trust violation by the authorization plane to enforcement of the autonomy restriction.

\textbf{Ground Truth.} Safety labels are determined independently of authorization decisions, based on authoritative transaction records, committed state versions, and disruption logs from the evaluation harness. The authorization plane relies solely on runtime-observable evidence, including state-version responses, authentication results, event metadata, and quorum status. It does not receive ground-truth labels or fault-injection schedules. An action is labeled unsafe when executing it would violate authoritative constraints, not merely because a disruption was injected. These independent labels distinguish unsafe executions from justified restrictions and unnecessary holds.

\textbf{Statistical Protocol.} Metrics are aggregated over 20 independent trials per scenario. Trial-level 95\% confidence intervals are reported for the primary S5 UAER comparison, and significance is assessed using Welch's $t$-test. Other tabulated metrics are descriptive, without reported confidence intervals or significance tests. Agent-specific metrics are reported only for configurations that support autonomous execution.
\section{Experimental Results}
\label{sec:results}

\subsection{Comparative Performance and Scenario Resilience}

\textbf{E1: Comparative Performance Across Baselines.} Table~\ref{tab:results} (Panel A) compares five configurations under compound stress (S5). GATE-DT reduced UAER from 8.8\% for the ungated agentic baseline B3 to 1.9\%, a 78.4\% reduction (95\% CI: 8.2--9.4\% vs.\ 1.6--2.2\% across 20 trials; Welch's $t(27.9) = 21.5$, $p < 0.001$). Stale-state execution, fraud acceptance, overselling, and duplicate side effects decreased by 67--85\% compared to B3. These gains reflect the combined execution-control architecture. Idempotency enforcement and version-bound checks may contribute to the reductions in duplicates and overselling. Differences among B0, B2, and B3 are reported but not causally isolated.

These improvements have trade-offs. Compared to B3, GATE-DT increased p95 authorization latency by 19~ms (92 vs.\ 73~ms), reduced order completion by 1.9 percentage points, and increased legitimate holds from 0.9\% to 4.1\%.

\textbf{E2: Fault and Attack Resilience.}
Table~\ref{tab:results} (Panel B) summarizes results across six scenarios (S0--S5). GATE-DT consistently recorded lower UAER than the ungated agentic baseline B3, with values from 0.2\% to 1.9\%, compared to 0.8\% to 8.8\% for B3. GATE-DT's legitimate order completion ranged from 98.8\% under normal operation (S0) to 94.2\% under compound disruptions (S5). Scenario-level hold rates and B3 completion rates were not reported. This limits comparison of operational continuity across configurations.

\textbf{Evidence Detection.}
Detection performance was evaluated against harness ground truth. For stale, reordered, or inconsistent CODT states, $R$ and $F$ scoring yielded precision and recall of 99.1\% and 98.7\% under S1, and 98.3\% and 97.6\% under S5, with mean injection-to-flag latencies of 14.2~ms and 18.4~ms, respectively. Under S3, $C$ assessment and hard vetoes blocked 98.0\% of replayed and 97.5\% of forged events. Additional authorization and execution controls held or denied a further 1.5\% and 2.0\%, respectively, resulting in a false acceptance rate of 0.5\% for both attack types, with a mean injection-to-deny time of 12.6~ms.

\begin{table}[t]
\caption{Comparative Performance and Scenario Resilience}
\label{tab:results}
\centering
\footnotesize
\setlength{\tabcolsep}{2pt}
\renewcommand{\arraystretch}{1.12}
\arrayrulecolor{black}
\setlength{\arrayrulewidth}{0.8pt}

\begin{tabular*}{\columnwidth}{@{\extracolsep{\fill}}
>{\columncolor{colorgray}}l
>{\columncolor{white}}c
>{\columncolor{white}}c
>{\columncolor{white}}c
>{\columncolor{white}}c
>{\columncolor{white}}c
@{}}
\hline
\multicolumn{6}{@{}l}{\textbf{Panel A: Baseline Comparison (S5)}} \\
\hline
\textbf{Metric} & \textbf{B0} & \textbf{B1} & \textbf{B2} & \textbf{B3} & \textbf{GATE-DT} \\
\hline
UAER (\%) & -- & -- & -- & 8.8 & \textbf{1.9} \\
Stale execution (\%) & 10.8 & 6.4 & 5.6 & 5.9 & \textbf{0.9} \\
Duplicate effects (\%) & 1.6 & 0.8 & 0.5 & 0.6 & \textbf{0.2} \\
Fraud acceptance (\%) & 7.4 & 6.9 & 2.2 & 5.3 & \textbf{1.4} \\
Overselling (\%) & 4.9 & 3.2 & 2.8 & 3.0 & \textbf{0.8} \\
p95 decision latency (ms) & \textbf{22} & 34 & 51 & 73 & 92 \\
Order completion (\%) & 92.0 & 94.8 & 93.5 & \textbf{96.1} & 94.2 \\
Legitimate holds (\%) & \textbf{0.4} & 0.7 & 2.1 & 0.9 & 4.1 \\
\hline
\multicolumn{6}{@{}l}{\textbf{Panel B: Fault and Attack Resilience}} \\
\hline
\textbf{Scenario} &
\shortstack{\textbf{B3}\\\textbf{UAER}} &
\shortstack{\textbf{GATE}\\\textbf{UAER}} &
\textbf{Stale} &
\textbf{Oversell} &
\textbf{Completion} \\
\hline
S0 & 0.8 & \textbf{0.2} & 0.1 & 0.1 & 98.8 \\
S1 & 5.1 & \textbf{1.0} & 0.5 & 0.4 & 96.8 \\
S2 & 6.4 & \textbf{1.3} & 0.6 & 0.5 & 95.7 \\
S3 & 7.3 & \textbf{1.4} & 0.3 & 0.4 & 96.0 \\
S4 & 5.9 & \textbf{1.1} & 0.6 & 0.7 & 95.0 \\
S5 & 8.8 & \textbf{1.9} & 0.9 & 0.8 & 94.2 \\
\hline
\end{tabular*}
\vspace{2pt}
\parbox{\columnwidth}{\scriptsize\textit{Note:} Panel B values are percentages. Stale execution, overselling, and completion refer to GATE-DT.}
\end{table}

\subsection{Trust Policy Sensitivity and Operational Trade-offs}

\begin{figure}[t]
\centering
\includegraphics[width=\columnwidth]{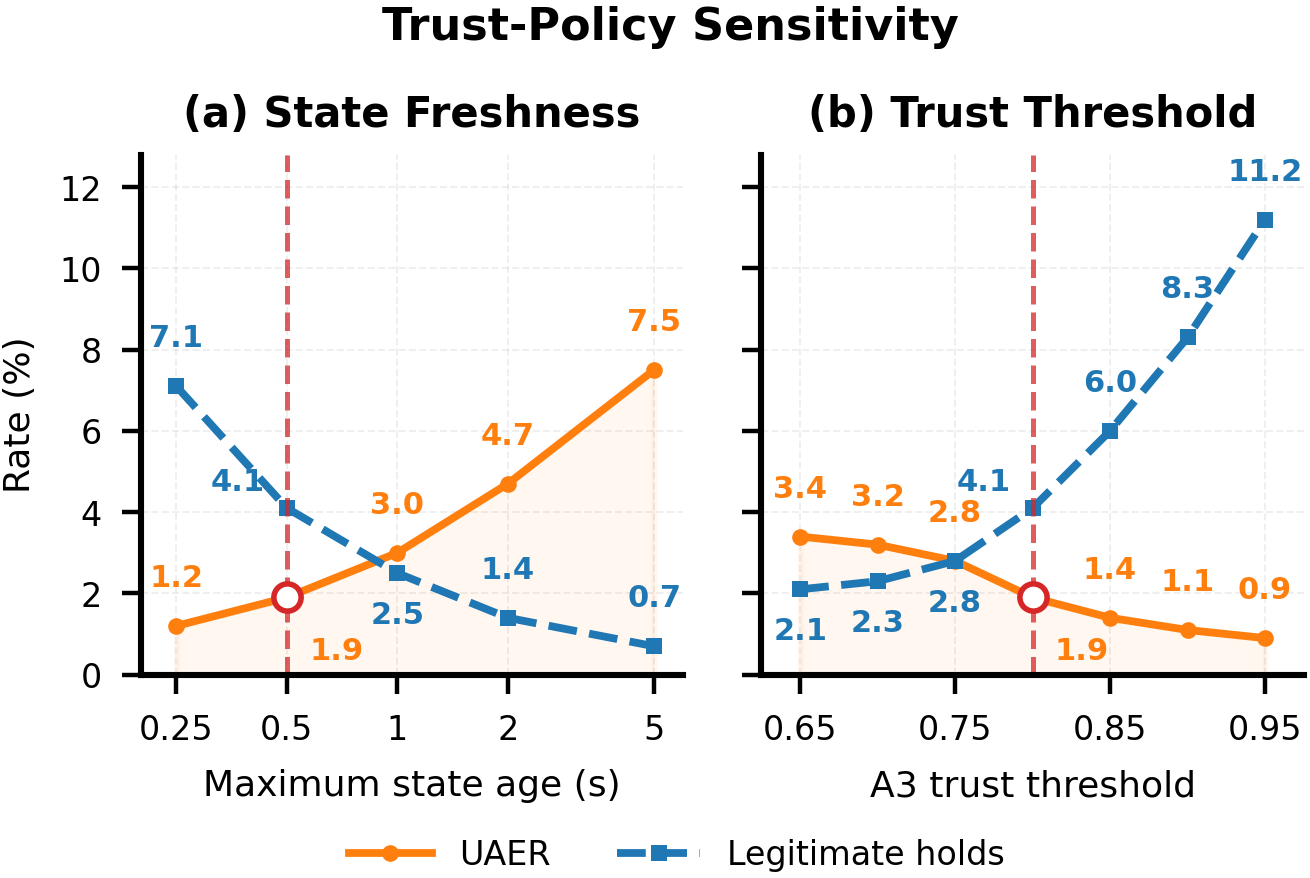}
\caption{Trust-policy sensitivity of GATE-DT under compound disruption (S5). (a) Maximum critical-state age varied from 0.25 to 5~s. (b) A3 trust threshold $\tau_a$ varied from 0.65 to 0.95. Red dashed lines and open markers indicate the reference policy (0.5~s, $\tau_a = 0.80$). Lower UAER is obtained at the cost of higher legitimate holds.}
\label{fig:policy_sensitivity}
\end{figure}

\textbf{E3: State-Freshness Sensitivity.} Fig.~\ref{fig:policy_sensitivity}(a) examines the effect of maximum critical-state age on execution safety and legitimate transaction holds. The maximum-age parameter modifies only the freshness hard-veto threshold, while the distributed-reliability scoring function $R$ and all other trust-policy parameters remain unchanged. With a strict freshness limit of 0.25~s, GATE-DT recorded a UAER of 1.2\% and a legitimate hold rate of 7.1\%. At the reference-policy limit of 0.5~s, UAER was 1.9\% and legitimate holds were 4.1\%. At 5.0~s, UAER increased to 7.5\% and legitimate holds dropped to 0.7\%. These results demonstrate that stricter freshness requirements improve execution safety, but increase transaction holds.

\textbf{E4: Trust-Threshold Sensitivity.} Fig.~\ref{fig:policy_sensitivity}(b) evaluates the effect of varying the A3 trust threshold $\tau_a$ from 0.65 to 0.95, with other controls held constant. UAER dropped from 3.4\% to 0.9\%, while legitimate holds increased from 2.1\% to 11.2\%. Below 0.75, UAER changed little, as dimension minimums and hard vetoes limited execution. Increasing $\tau_a$ from 0.75 to 0.80 reduced UAER by 0.9 percentage points (pp) with a 1.3-pp increase in holds. Increasing it from 0.80 to 0.85 reduced UAER by only 0.5~pp with a 1.9-pp increase. Thus, a threshold of 0.80 lies at the knee of the trade-off. Scaling all dimension minimums $\delta_j$ from 0.8$\times$ to 1.2$\times$ at $\tau_a = 0.80$ showed the same pattern. UAER decreased from 2.7\% to 1.4\% and legitimate holds increased from 2.9\% to 7.8\%, with the reference scaling (1.0$\times$) also at the knee.

\subsection{Component Ablation Analysis}

\textbf{E5: Trust Component Ablation.}
Table~\ref{tab:ablation} (Panel A) examines selected trust-evidence and authorization components by disabling them individually under compound stress (S5). The complete GATE-DT configuration recorded a UAER of 1.9\%. Removing the hard authority and veto gate produced the largest increase, raising UAER to 8.4\%. Disabling the freshness hard gate increased UAER to 7.8\%. This ablation removes only the age-based hard veto, retaining freshness assessment within the distributed-reliability score $R$ and all other authorization controls. Removing cyber integrity assessment raised UAER to 6.8\%, while removing distributed reliability assessment increased it to 5.4\%. Excluding operational fidelity and provenance assessment resulted in rates of 2.9\% and 2.6\%, respectively.

Removing data quality ($D$) and model confidence ($U$) increased UAER to 2.2\% and 2.0\%, respectively. The limited impact of $U$ aligns with its role as uncalibrated, agent-reported evidence. Replacing adaptive autonomy with a fixed A3 policy increased UAER to 2.5\%, while disabling execution-time revalidation increased it to 3.6\%. The gate-removed configuration records 8.4\% UAER, compared to 8.8\% for B3. Since this configuration retains other execution safeguards, the 0.4-pp difference cannot be attributed to a specific mechanism.

\textbf{S5 Disruption Contribution.}
Table~\ref{tab:ablation} (Panel B) removes each disruption from S5 individually. Removing replication lag narrowed the B3 to GATE-DT gap the most, from 6.9~pp to 5.2~pp, followed by infrastructure failure (5.6~pp), demand surge (5.8~pp), replay attacks (5.9~pp), and fraudulent requests (6.1~pp). The gap remained at least 5.2~pp in all cases, indicating that no single disruption explains the improvement. The individual reductions total 5.9~pp, suggesting that interactions among concurrent disruptions account for the remainder.

\subsection{Authority Degradation and Recovery}

\textbf{E6: Authority Degradation and Recovery.} Table~\ref{tab:ablation} (Panel C) presents authority downgrade and recovery times across five disruption scenarios. Excluding detection delay, GATE-DT enforced authority downgrades within 24--61~ms, fastest under cyber and fraud attacks (S3). Following disruption removal, recovery to A3 took 56--118~s, longest under compound disruptions (S5). The effects of intermediate transition thresholds and evidence-hold durations were not separately assessed.

\begin{table}[t]
\caption{Trust Component Ablation, S5 Disruption Contribution, and Authority Recovery}
\label{tab:ablation}
\centering
\footnotesize
\setlength{\tabcolsep}{3pt}
\renewcommand{\arraystretch}{1.12}
\arrayrulecolor{black}
\setlength{\arrayrulewidth}{0.8pt}

\begin{tabular}{@{}
>{\columncolor{colorgray}}l
>{\columncolor{white}}c
>{\columncolor{colorgray}}c
@{}}
\hline
\multicolumn{3}{@{}l}{\textbf{Panel A: Trust Component Ablation (S5)}} \\
\hline
\textbf{Configuration} & \textbf{UAER (\%)} & \textbf{$\Delta$ UAER (pp)} \\
\hline
Full GATE-DT & \textbf{1.9} & -- \\
Without freshness hard gate & 7.8 & +5.9 \\
Without distributed reliability ($R$) & 5.4 & +3.5 \\
Without cyber integrity ($C$) & 6.8 & +4.9 \\
Without operational fidelity ($F$) & 2.9 & +1.0 \\
Without provenance ($P$) & 2.6 & +0.7 \\
Without data quality ($D$) & 2.2 & +0.3 \\
Without model confidence ($U$) & 2.0 & +0.1 \\
Fixed A3 (no adaptive autonomy) & 2.5 & +0.6 \\
Without execution-time revalidation & 3.6 & +1.7 \\
Without authority/veto gate & 8.4 & +6.5 \\
\hline
\multicolumn{3}{@{}l}{\textbf{Panel B: S5 Leave-One-Out UAER}} \\
\hline
\textbf{Removed Disruption} & \textbf{B3/GATE (\%)} & \textbf{Gap (pp)} \\
\hline
None (full S5) & 8.8/\textbf{1.9} & 6.9 \\
Replication lag & 6.7/1.5 & 5.2 \\
Infrastructure failure & 7.2/1.6 & 5.6 \\
Replay attacks & 7.6/1.7 & 5.9 \\
Fraudulent requests & 7.8/1.7 & 6.1 \\
Demand surge & 7.4/1.6 & 5.8 \\
\hline
\multicolumn{3}{@{}l}{\textbf{Panel C: Authority Degradation and Recovery}} \\
\hline
\textbf{Scenario} & \textbf{Downgrade (ms)} & \textbf{Recovery (s)} \\
\hline
S1: State degradation & 32 & 56 \\
S2: Infrastructure failure & 45 & 69 \\
S3: Cyber/fraud attack & 24 & 82 \\
S4: Demand surge & 39 & 64 \\
S5: Compound disruption & 61 & 118 \\
\hline
\end{tabular}
\end{table}
\section{Discussion}
\label{sec:discussion}

\textbf{Separating Decision and Execution Authority.} The results demonstrate the value of separating autonomous decision generation from execution authorization. Ablation analysis highlights the contributions of hard authority enforcement, freshness enforcement, cybersecurity integrity, distributed-state reliability, and execution-time revalidation to execution safety. This approach aligns with action shielding, which prevents unsafe autonomous actions before execution \cite{alshiekh2018shielding}, and extends protection to distributed transactional state and cybersecurity evidence.

\textbf{Safety-Continuity Trade-off.} UAER measures only executed autonomous actions, so its reduction must be interpreted alongside legitimate holds and order completion. Stricter freshness limits and higher trust thresholds reduce unsafe executions but increase legitimate holds, reflecting a trade-off analogous to risk-coverage in selective prediction \cite{geifman2019selectivenet}. Under S5, GATE-DT incurred 4.1\% legitimate holds and a 1.9-pp reduction in completion compared to B3. Across 20 trials, GATE-DT prevented approximately 2.2 unsafe autonomous executions for each additional legitimate hold. Adaptive autonomy manages this trade-off by restricting execution authority when trust deteriorates and restoring it incrementally as trust improves.

\textbf{Limitations and Future Work.} Results come from an emulation harness with six predefined scenarios, a fixed reference configuration, and fixed trust weights. This limits generalizability to production environments. The minimum thresholds $\delta_j$ are fixed reference-policy constraints, evaluated only through uniform scaling. The agent-reported confidence score $U$ has not been independently calibrated. Security guarantees depend on the stated trust assumptions for the authorization plane. The business costs of legitimate holds were not quantified. Future work should investigate broader deployments and threat scenarios, adaptive trust calibration, per-dimension trust-score distributions, and the economic impact of safety-continuity trade-offs.
\section{Conclusion}
\label{sec:conclusion}

GATE-DT integrates operational digital twins, multidimensional trust assessment, and adaptive autonomy to govern transaction-sensitive actions in distributed retail and e-commerce systems. The central finding is that autonomous execution safety relies on both agent decisions and the reliability and security of the distributed state. Under compound stress, GATE-DT reduced untrustworthy autonomous execution from 8.8\% to 1.9\%, and also decreased stale-state execution, fraud acceptance, and overselling. Ablation results highlight that explicit execution authority, freshness enforcement, cybersecurity integrity, distributed reliability, and execution-time revalidation are critical. These safety improvements introduce additional authorization latency and legitimate transaction holds, underscoring the trade-off between protection and operational continuity. The results suggest treating execution authority as a dynamic, evidence-based privilege rather than a permanent capability.

\balance
\bibliographystyle{IEEEtran}
\bibliography{references}

\end{document}